\documentclass[11pt,a4paper]{article}
\usepackage{jcappub,graphics,subfigure}

\newcommand\e{\mathrm{e}}
\newcommand\ele{\mathrm{l}}
\newcommand\ere{\mathrm{r}}

\newcommand\im{\mathrm{i}}

\newcommand\num{\mathrm{num}}
\newcommand\phai{\mathrm{pi}}
\newcommand\sr{\mathrm{sr}}
\newcommand\real{\mathrm{Re}}
\newcommand\ret{\mathrm{ret}}
\newcommand\s{\mathrm{s}}
\newcommand\sca{\mathrm{S}}

\newcommand\ua{\mathrm{ua}}

\newcommand\D{\mathrm{d}}

\newcommand\Mpc{\mathrm{Mpc}}
\newcommand\Pl{\mathrm{Pl}}

\title{Computation of the Power Spectrum in Chaotic $\frac{1}{4}\lambda\phi^4$ Inflation}

\author[a]{Clara Rojas} \author[b]{and V\'ictor M. Villalba} 

\affiliation[a]{Instituto Venezolano de Investigaciones Cient\'ificas (IVIC), Centro de Estudios Interdisciplinarios de la F\'isica, Caracas 1020A, Venezuela.} \affiliation[b]{School of Mathematical Sciences, Faculty of Science, Monash University, Australia}

\emailAdd{clararoj@gmail.com} \emailAdd{Victor.Villalba@monash.edu} 

\abstract{The phase-integral approximation devised by Fr\"oman and Fr\"oman, is used for computing cosmological  perturbations in the quartic chaotic inflationary model. The phase-integral formulas for the scalar power spectrum are explicitly obtained up to fifth order of the phase-integral approximation. As in  previous reports \cite{rojas:2007b,rojas:2007c,rojas:2009}, we point out that the accuracy of the phase-integral approximation compares favorably with the numerical results and those obtained using the slow-roll  and uniform approximation methods.}


\begin{document}
\maketitle

%************************************************************************************
\section{Introduction}
%************************************************************************************

Chaotic inflationary models like $V \propto \phi^\mathrm{p}$ have been studied and analized with the WMAP data in recent years. The recent observations reported by WMAP favors inflation,  and the quartic chaotic model, with $\mathrm{p}=4$ \cite{linde:1983a,linde:1983b}  still remains ruled out by this data  \cite{komatsu:2011,komatsu:2009}.  Ramirez {\it et al.} \cite{ramirez:2009} point out that  if  the scalar power spectrum is considered not scale invariant,  then quartic models would not be excluded by the recent cosmological data.  Ramirez {\it et al.} \cite{ramirez:2009}  consider a new scenario where the observable modes cross the horizon near the beginning of inflation.

In order to compare with observations,  we should be able to obtain very accurate results for the predicted power spectrum of primordial perturbations for a variety of inflationary scenarios. In general, most of the inflationary models are not exactly solvable and approximate or numerical methods are mandatory in the computation of the scalar and power spectra.  Traditionally, the method of approximation applied in inflationary cosmology is the slow-roll approximation \cite{stewart:1993}, which produces reliable results in inflationary models with smooth potentials,  but cannot
be improved  on a simple way beyond the leading order. Recently, some authors have applied alternative approximations, such as the WKB method with the Langer modification \cite{langer:1937,martin:2003a,casadio:2005a}, the Green function method \cite{stewart:2001}, and the improved WKB method \cite{casadio:2005c,casadio:2005b}.

Habib \textit{et al} \cite{habib:2002,habib:2004,habib:2005b}  have successfully applied the uniform approximation method in the calculation of the scalar and tensor power spectra and the corresponding spectral indices for the quadratic and quartic chaotic inflationary models, showing that the the uniform approximation gives more accurate results than the slow-roll approximation. Casadio \textit{et al} \cite{casadio:2006} have applied  the method of comparison equation to study cosmological perturbations during inflation.  The comparison method is based on  the uniform approximation proposed by Dingle \cite{dingle:1956} and Miller \cite{miller:1953} and thoroughly discussed by Berry and Mount \cite{berry:1990}.

Recently \cite{rojas:2007b,rojas:2007c,rojas:2009}, the phase integral method has been used in the calculation of the scalar and tensor power spectra and the spectral indexes for the power-law and quadratic chaotic model, obtaining better results than  those derived using slow-roll \cite{stewart:1993}, WKB and the improved uniform approximation methods \cite{habib:2002,habib:2004,habib:2005b}.  

Since the 7-year WMAP data does not show evidence for tensor modes \cite{larson:2011,alabidi:2010}, in this article we use the phase-integral approximation up to fifth order to calculate the scalar power spectrum for the quartic chaotic inflationary model. We have calculated the tensor power spectrum and  since the results  are very similar to those obtained in the scalar case, we decided to skip their presentation in this article.f

The article is structured as follows: In Sec. \ref{perturbations} we numerically solve the equation governing the scalar perturbations. In Sec. \ref{methods} we apply the phase-integral approximation to the quartic chaotic inflationary model.  We present the improved uniform approximation method,  the slow-roll approximation and the numerical solution to the perturbation equation. The comparison among different  methods is presented in Sec. \ref{results}. Finally in Sec. \ref{conclusion} we present our conclusions.

%************************************************************************************
\section{Equation for the perturbations}
\label{perturbations}
%************************************************************************************

In an inflationary universe driven by a scalar field, the equations of motion for the inflaton $\phi$ and the Hubble parameter $H$ are given by \cite{liddle:2000b}

\begin{eqnarray}
\label{ddotphi}
\ddot{\phi}&+&3H\dot{\phi}=-\frac{\partial V(\phi)}{\partial\phi},\\
\label{H^2}
H^2&=&\frac{1}{3M_\Pl^2}\left[V(\phi)+\frac{1}{2}\dot{\phi}^2\right],
\end{eqnarray}
with $V(\phi)=\frac{1}{4}\lambda\phi^4$ and the constant  $\lambda$, can be fixed with the help of the amplitude of the density fluctuations observed by WMAP. A fitting to the observational data requires that  $\lambda\simeq 10^{-13}$ \cite{habib:2005b}.
Using Eq. \eqref{ddotphi} and Eq. \eqref{H^2} we obtain that, in the slow-roll approximation, the expansion factor $a(t)$ and the inflaton field $\phi(t)$ are

\begin{eqnarray}
\label{asr}
a&\simeq& a_i \left\{\frac{\phi_i^4}{8\,M_\Pl^2}\left[1-\exp\left(-\sqrt{\frac{16\lambda}{3}}
\,M_\Pl t\right)\right]\right\},\\
\label{phisr}
\phi(t)&\simeq&\phi_i\exp\left(-2\sqrt{\frac{\lambda}{3}} \,M_\Pl t\right),
\end{eqnarray}
where $a_i$ and $\phi_i$ are integration constants corresponding to the initial values of the scale factor and the inflaton. 

The application of the the phase-integral approximation for solving the equations governing the scalar perturbations requires the knowledge of the evolution of the expansion factor $a$ and the scalar field $\phi$, and the computation of analytic expression of them would highly simplify the calculations. In order to find $\phi$ and $a$ as a function of  time, we solve the coupled system of differential equations (\ref{ddotphi})-(\ref{H^2}) with the sixth-order Runge-Kutta method \cite{gerald:1984} and fit the numerical data with the software {\it gnuplot} following the form of the slow-roll solutions (\ref{asr},\ref{phisr}). The integration is performed in the physical time $t$ using the sixth-order Runge-Kutta method \cite{gerald:1984}. The initial value for the inflaton is $\phi_i=21.9 \,M_\Pl$. The initial condition for the speed of the scalar field  $\dot{\phi}_i$ is obtained from the slow-roll solution (\ref{phisr}).  The initial value for  $a_i$ is  chosen as unity and   $\lambda=1.75\times 10^{-13}$. These initial conditions guarantee sufficient  inflation.  We find that the inflation ends at $t_f=4.62\times 10^6 \,M_\Pl^{-1}=1.25\times 10^{-36}\s$, corresponding to $59.98$ e-folds before the scalar field starts to oscillate.

The equation governing the evolution of the scalar perturbations in the quartic chaotic inflationary model is given by 

\begin{eqnarray}
\label{dotu}
\ddot{u_k}+\frac{\dot{a}}{a}\dot{u_k}+\frac{1}{a^2}\left[k^2-\frac{\left(\dot{a}\dot{z_\sca}+a\ddot{z_\sca}\right)a}{z_\sca} \right]u_k&=&0.
\end{eqnarray}

In order to apply the phase-integral approximation  we need the expressions for $a(t)$ and  $\phi(t)$ which have been derived  fitting  the numerically solution of  Eqs.  (\ref{ddotphi}) and (\ref{H^2}).  We have obtained the following analytic expressions

\begin{eqnarray}
\label{ch4_a_fit}
a(t)&=&a_i\exp\left\{\bar{\alpha} \left[1-\exp\left(-\bar{\beta} t\right)\right]\right\},\\
\label{ch4_phi_fit}
\phi(t)&=&\phi_i\exp\left(-\bar{\gamma} t\right),
\end{eqnarray}
where $\bar{\alpha}=60.2839$,
$\bar{\beta}=9.6370\times10^{-7}\,M_\Pl$, and $\bar{\gamma}=4.8305\times10^{-7}\,M_\Pl$ and, the function $z_\sca(t)$ has he form

\begin{equation}
\label{ch4_zs}
z_\sca(t)=-a_i\phi_i\frac{\bar{\gamma}}{\bar{\alpha}\bar{\beta}} \exp\left\{\bar{\alpha}\left[1-\exp\left(-\bar{\beta} t\right)\right]+\bar{\beta} t-\bar{\gamma} t\right\}.
\end{equation}

Since we  want  to eliminate the term $\dot{u}_k$ in Eq. (\ref{dotu}),  we introduce the change of variable $u_k(t)=\frac{U_k(t)}{\sqrt{a}}$, obtaining that $U_{k}$ satisfies the differential equation:

\begin{eqnarray}
\label{ch4_ddotUk}
\ddot{U}_k+R_\sca(k,t)U_k&=&0,
\end{eqnarray}
with

\begin{eqnarray}
\label{ch4_RS0}
R_\sca(k,t)&=&\frac{1}{a^2}\left[k^2-\frac{\left(\dot{a}\dot{z_\sca}+a\ddot{z_\sca}\right)a}{z_\sca} \right]+\frac{1}{4a^2}\left(a^2-2a\ddot{a}\right),
\end{eqnarray}
where $U(k)$ satisfies the asymptotic conditions

\begin{eqnarray}
\label{ch4_cero_Uk}
U_k&\rightarrow&A_k \sqrt{a(t)} z_\sca(t),\quad  k\,t\rightarrow \infty,\\
\label{ch4_borde_Uk}
U_k&\rightarrow&\sqrt{\frac{a(t)}{2k}}\exp{\left[-ik\eta(t)\right]}, \quad k\,t\rightarrow 0.
\end{eqnarray}

We now proceed to write the explicit equations for quartic chaotic inflation. From  Eq. (\ref{ch4_RS0}),  with Eq. (\ref{ch4_a_fit}) and  Eq. (\ref{ch4_phi_fit}) we obtain

\begin{eqnarray}
\label{ch4_RS}
R_\sca(k,t)&=& \frac{k^2}{a_i^2}\exp\left\{-2\bar{\alpha}\left[1-\exp\left(-\bar{\beta} t\right)\right]\right\}-\frac{9}{4}\bar{\alpha}^2\bar{\beta}^2\exp\left(-2\bar{\beta} t\right)\\
\nonumber
&-&\frac{3}{2}\bar{\alpha}\bar{\beta}\left(\bar{\beta}-2\bar{\gamma}\right)\exp\left(-\bar{\beta} t\right)-\left(\bar{\beta}-\bar{\gamma}\right)^2.
\end{eqnarray}

\begin{figure}[htbp]
\vspace{1cm}
\begin{center}
\includegraphics[scale=0.50]{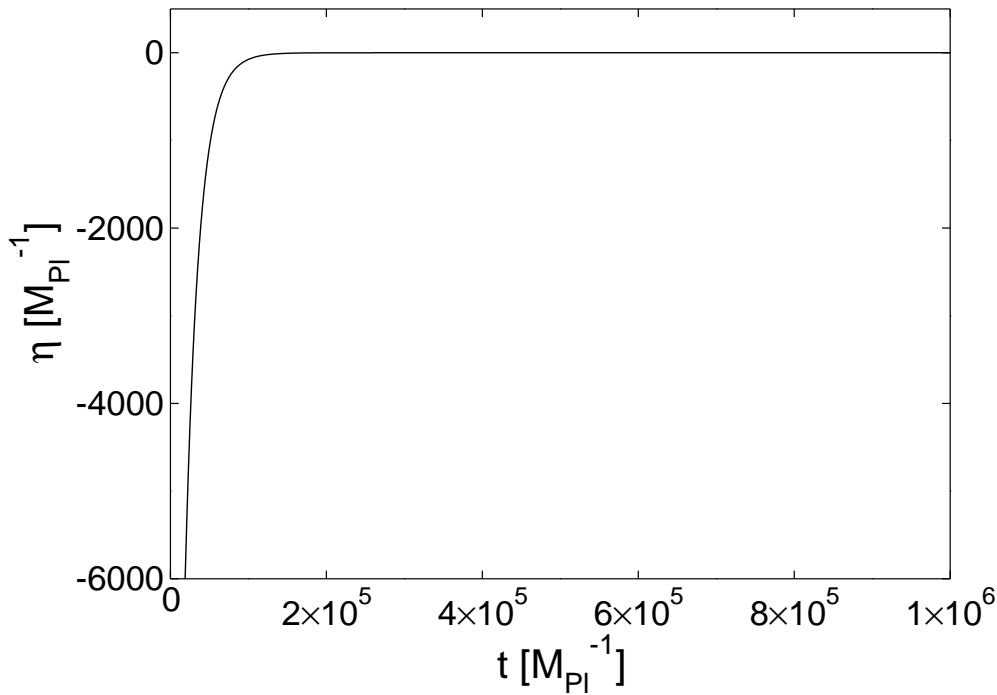}
\caption{\small{Behavior of  $\eta$ as a function of $t$ for the chaotic inflationary $\frac{1}{4}\lambda \phi^4$ model.}}
\label{ch4:eta}
\end{center}
\end{figure}

In order to apply  the asymptotic condition (\ref{ch4_borde_Uk}), we  use the relationship  between  $\eta$ and the physical time  $t$,  which is given in terms of the integral exponential function $\mathrm{Ei}(z)$ \cite{abramowitz:1965} as

\begin{equation}
\eta(t)=\frac{1}{a_i\bar{\beta}} \exp\left(-\bar{\alpha}\right)\left\{\mathrm{Ei}\left[\bar{\alpha} \exp\left(-\bar{\beta} t_0\right)\right]-\mathrm{Ei}\left[\bar{\alpha} \exp\left(-\bar{\beta} t\right)\right]\right\},
\end{equation}
where  $t_i=10^7 \,M_\Pl^{-1}$,  so that  $\eta$ is zero at the end of the inflationary epoch. The dependence  of $\eta$ on $t$ is shown in Fig. \ref{ch4:eta}. We can observe that,  as

\begin{eqnarray}
-k\,\eta\rightarrow 0       &\Rightarrow&  k\,t \rightarrow \infty,\\
-k\,\eta\rightarrow \infty  &\Rightarrow& k\,t \rightarrow 0.
\end{eqnarray}

Eq. (\ref{ch4_ddotUk}), where $R_\sca(k,t)$ is given by Eq. (\ref{ch4_RS}) , does not  possess exact analytic solution. In order to solve the differential equations governing the scalar perturbations in the physical time $t$,  we use the fifth-order phase integral approximation \cite{froman:1996} and compare this results with  those obtained  using the slow-roll and the improved uniform approximation. 

Once the mode equation for scalar perturbations is solved for different momenta $k$, the power spectra for scalar modes is given by the expression

\bigskip
 \begin{eqnarray}
 \label{PS}
 P_\sca(k)&=& \lim_{-k\eta\rightarrow 0} \frac{k^3}{2 \pi^2}\left|\frac{u_k(\eta)}{z(\eta)} \right|^2.
 \end{eqnarray}

\vspace{1cm}
The spectral indice is defined as

\begin{eqnarray}
 n_\sca(k)&=&1+\frac{d \ln P_\sca(k)(k)}{d \ln k}.
\end{eqnarray}

%************************************************************************************
\section{Methodology}
\label{methods}
%************************************************************************************

\subsection{Phase-integral approximation}

\begin{figure}[htbp]
\begin{center}
\subfigure[]{
\label{ch4:QSa}
\includegraphics[scale=0.45]{ch4_QS.eps}}
\subfigure[]{
\label{ch4:QSb}
\includegraphics[scale=0.40]{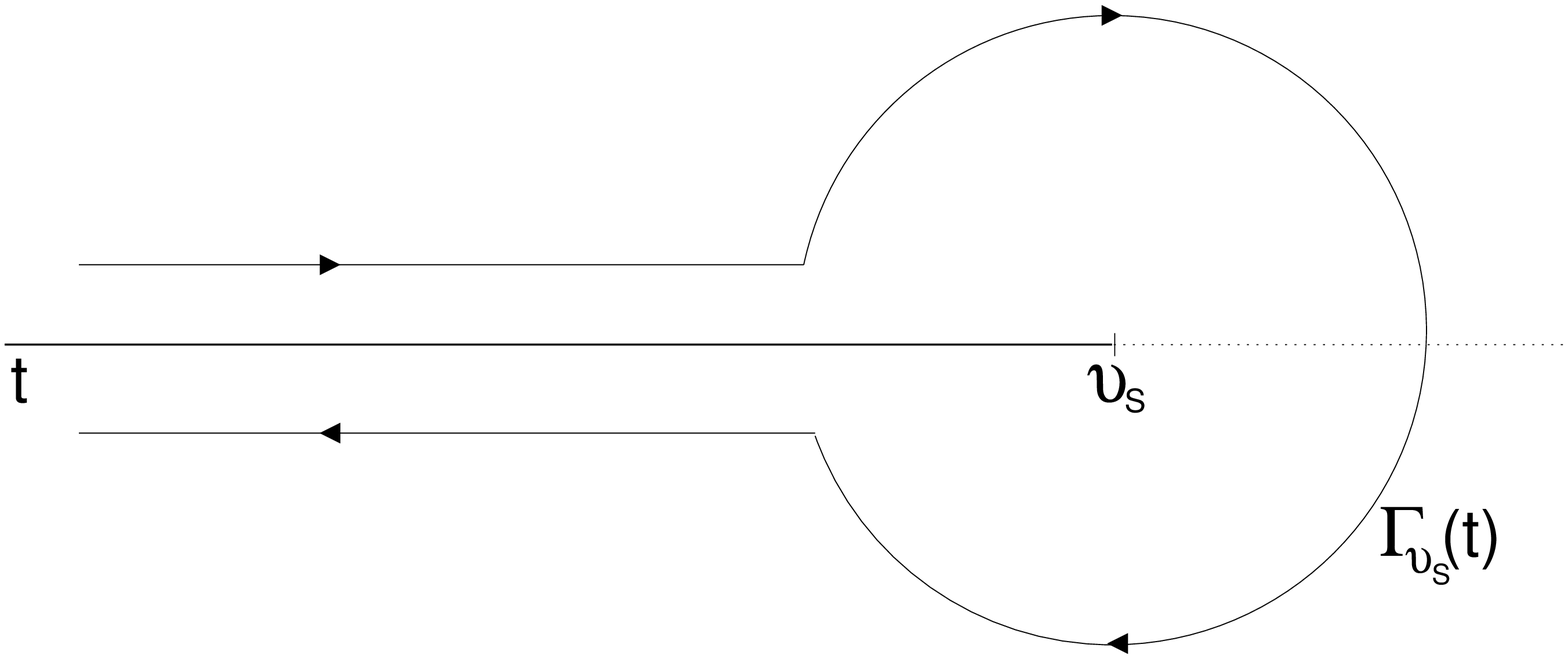}}
\subfigure[]{
\label{ch4:QSc}
\includegraphics[scale=0.40]{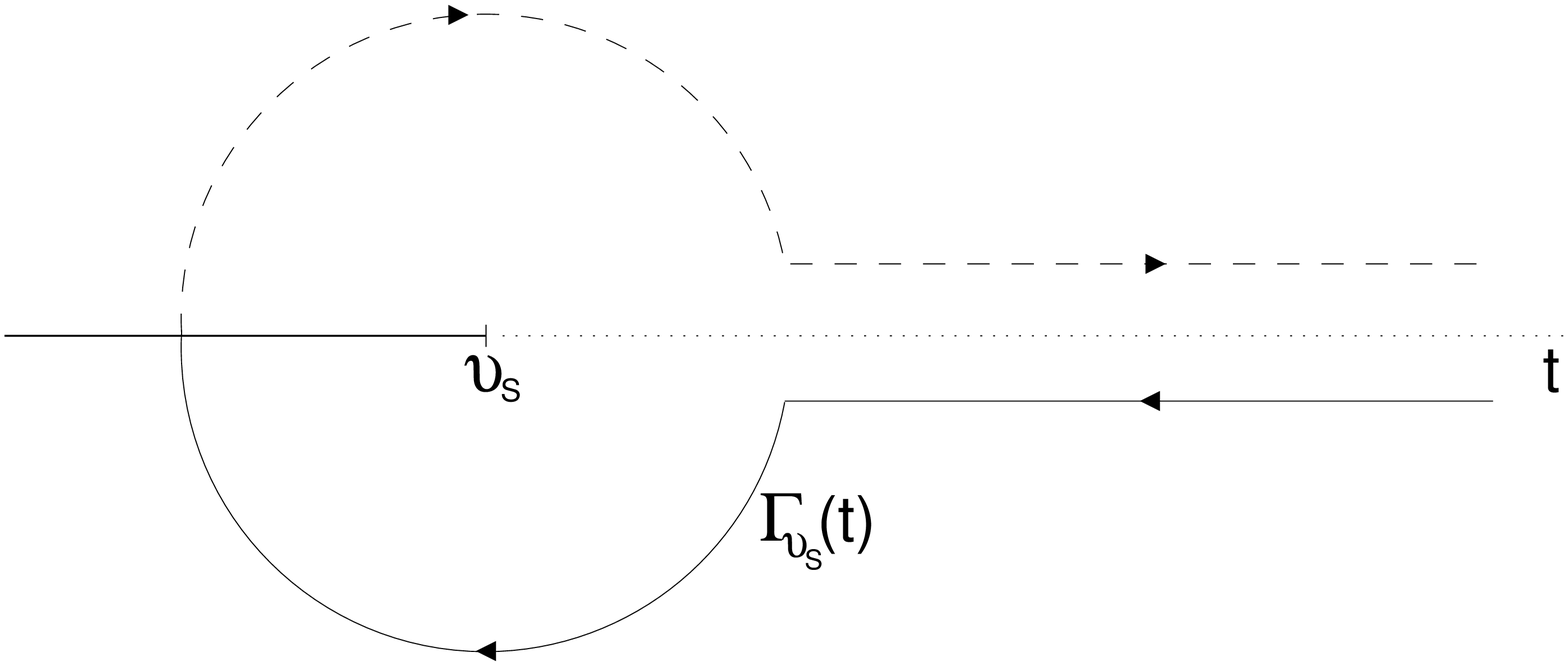}}
\caption{\small{(a) Behavior of  $Q_\sca^2(k,t)$.
(b) Contour of integration $\Gamma_{\nu_\sca}(t)$ for $0<t<\nu_\sca$. 
(c) Contour of integration  $\Gamma_{\nu_\sca}(t)$ for $t>\nu_\sca$. 
The contour of integration has been done in the complex $z$ plane. For this function we have a Riemann surface with two sheets where the function is single-valued. In the contour of integration the solid line indicates the first Riemman sheet whereas that the second Riemman sheeet is indicated for the dashed line \cite{froman:1986}.}}
\end{center}
\end{figure}

In order to solve Eq. (\ref{ch4_ddotUk})  with the help of the phase-integral approximation \cite{froman:1996}, we choose the following base functions $Q_\sca$ for the scalar  perturbations

\begin{eqnarray}
\label{Q}
Q_\sca^2(k,t)&=&R_\sca(k,t),
\end{eqnarray}
where  $R_\sca(k,t)$ is given by  Eq. (\ref{ch4_RS}).  Using this selection, the phase-integral approximation is  valid as  $k t\rightarrow \infty$, limit where we should impose the condition (\ref{ch4_cero_Uk}), where the validity condition   $\mu \ll 1$ holds. The selection, given in Eq. (\ref{Q}), makes the first order phase-integral approximation coincides with the WKB solution. The bases functions  $Q_\sca(k,t)$  possess turning points  $t_\ret=\upsilon_\sca=117726\,M_\Pl^{-1}$ for the mode $k=0.05\,\Mpc^{-1}$. The turning point represents the horizon. There are two ranges  where to define the solution. To the left of the turning point  $0<t<t_\ret$ we have the classically permitted region  $Q_{\sca}^2(k,t)>0$ and to the right of the turning point $t>t_\ret$ corresponding to the classically forbidden region $Q_{\sca}^2(k,t)<0$, such as it is shown in Figs \ref{ch4:QSa}.

The mode  $k$ equations for the scalar  perturbations (\ref{ch4_ddotUk})  in the phase-integral approximation has two solutions:  For $0<t <t_\ret$

\begin{eqnarray}
\label{ch4_uk_left}
u^\phai_k(t)&=& \frac{c_1}{\sqrt{a(t)}}\left|q_\sca^{-1/2}(k,t)\right| \cos{\left[\left|\omega_\sca(k,t)\right|-\frac{\pi}{4}\right]} \\
\nonumber
&+& \frac{c_2}{\sqrt{a(t)}}\left|q_\sca^{-1/2}(k,t)\right| \cos{\left[\left|\omega_\sca(k,t)\right|+\frac{\pi}{4}\right]}.
\end{eqnarray}

For $t>t_\ret$

\begin{eqnarray}
\label{ch4_uk_right}
u^\phai_k(t)&=&\frac{c_1}{2\sqrt{a(t)}}\left|q_\sca^{-1/2}(k,t)\right|\exp\left[-\left|\omega_\sca(k,t)\right|\right]\\
 \nonumber
&+& \frac{c_2}{\sqrt{a(t)}} \left|q_\sca^{-1/2}(k,t)\right| \exp\left[\left|\omega_\sca(k,t)\right|\right].
\end{eqnarray}

The phase-integral approximation solutions are of the order $2N+1$, so $N=2$ means that we have the approximation up to fifth order and  $q_{\sca}(k,t)$  can be expanded in the form

\begin{eqnarray}
\label{q1}
q_\sca(k,t)=\sum_{n=0}^2 Y_{2n_\sca}(k,t) Q_\sca(k,t)=\left[Y_{0_\sca}(k,t)+Y_{2_\sca}(k,t)+Y_{4_\sca}(k,t)\right] Q_\sca(k,t).
\end{eqnarray}

In order to compute $q_\sca(k,t)$, we  compute $Y_{2_\sca}(k,t)$, $Y_{4_\sca}(k,t)$, and the required functions  $\varepsilon_{0_\sca}(k,t)$, $\varepsilon_{2_\sca}(k,t)$.  The expression  (\ref{q1}) gives  a fifth-order approximation for $q_\sca(k,t)$.  In order to compute $\omega_\sca(k,t)$  we make a contour integration following the path indicated in Fig. \ref{ch4:QSb}-(c).

\begin{eqnarray}
\omega_\sca(k,t)&=&\omega_{0_\sca}(k,t)+\sum_{n=1}^{2} \omega_{2n_\sca}(k,t),\\
&=&\int_{\upsilon_\sca}^{t}Q_\sca(k,t)\D t+\frac{1}{2}\sum_{n=1}^2\int_{\Gamma_{\upsilon_\sca}}Y_{2n_\sca}(k,t)Q_\sca(k,t)\D t,\\
&=&\int_{\upsilon_\sca}^{t}Q_\sca(k,t)\D t+\frac{1}{2}\sum_{n=1}^2\int_{\Gamma_{\upsilon_\sca}}f_{2n_\sca}(k,t)\D t,
\end{eqnarray}
\medskip
where
\begin{eqnarray}
f_{2n_\sca}(k,t)&=&Y_{2n_\sca}(k,t)Q_\sca(k,t).
\end{eqnarray}
The functions $f_{2n_\sca}(k,t)$  have the following functional dependence:

\begin{eqnarray}
\label{A}
f_{2_\sca}(k,t)&=&A(k,t)(t-\upsilon_\sca)^{-5/2},\\
f_{4_\sca}(k,t)&=&B(k,t)(t-\upsilon_\sca)^{-11/2},\\
\end{eqnarray}
where $A(k,t)$ and $B(k,t)$ are regular at $\upsilon_\sca$ . With the help of  the functions  (\ref{A}) we compute the integrals for $\omega_{2n}$ up to $N=4$ using the contour indicated in  Fig. \ref{ch4:QSb}-(c). The expressions for  $\omega_{2n}$ permit one to obtain the fifth-order phase integral approximation of the solution to the equation for scalar perturbations (\ref{ch4_ddotUk}).  The constants $c_1$ and $c_2$ are obtained using the limit  $k\,t\rightarrow 0$ of the solutions on the left side of the turning point (\ref{ch4_uk_left}), and  are given by the expressions

\begin{eqnarray}
c_1&=&-\im\,c_2,\\
c_2&=&\frac{\e^{-\im\frac{\pi}{4}}}{\sqrt{2}}\e^{-\im\left[k\,\eta(0)+\left|\omega_{0_\sca}(k,0)\right|\right]}.
\end{eqnarray}
In order to compute the scalar power spectrum, we need to calculate the limit as  $k\,t\rightarrow \infty$ of the growing part of the solutions on the right side of the turning point  given by   Eq. (\ref{ch4_uk_right}) for scalar perturbations.

\begin{eqnarray}
P_\sca(k)&=&\lim_{-k t\rightarrow \infty} \frac{k^3}{2\pi^2} \left|\frac{u_k^\phai(t)}{z_\sca(t)}\right|^2.
\end{eqnarray}

%************************************************************************************
\subsection{Improved Uniform approximation}
%************************************************************************************

We want to obtain an approximate solution to the differential equation (\ref{ch4_ddotUk})  in the range where  $Q_\sca^2(k,t)$ have a simple root  at $t_\ret=\upsilon_\sca$, so that $Q_{\sca}^2(k,t)>0$ for  $0<t<t_\ret$ and $Q_{\sca}^2(k,t)<0$ for  $t>t_\ret$ as depicted in Fig. \ref{ch4:QSa}. Using the uniform approximation method \cite{berry:1972,habib:2002,rojas:2007b,rojas:2007c,rojas:2009}, we obtain that for  $0<t<t_\ret$ we have

\begin{eqnarray}
\label{Uk_zero}
U_k(k,t)&=&\left[\frac{\rho_\ele(k,t)}{Q_\sca^2(k,t)} \right]^{1/4} \left\{C_1
A_i[-\rho_\ele(k,t)]+C_2 B_i[-\rho_\ele(k,t)] \right\},\\
\frac{2}{3}\left[\rho_\ele(k,t)\right]^{3/2}&=&\int_{t}^{t_\ret} \left[Q_{\sca}^2(k,t)\right]^{1/2}\D t,
\end{eqnarray}
\medskip
where  $C_1$ and  $C_2$ are two constants to be determined with the help of the boundary conditions (\ref{ch4_borde_Uk}). For $t> t_\ret$

\begin{eqnarray}
\label{Uk_infinity}
U_k(k,t)&=&\left[\frac{-\rho_\ere(k,t)}{Q_\sca^2(k,t)} \right]^{1/4} \left\{C_1
A_i[\rho_\ere(k,t)]+C_2 B_i[\rho_\ere(k,t)] \right\},\\
\frac{2}{3}\left[\rho_\ere(k,t)\right]^{3/2}&=&\int_{t_\ret}^{t} \left[-Q_{\sca}^2(k,t)\right]^{1/2}\D t,
\end{eqnarray}

For the computation of the power spectrum we need to take the limit $k\,t\rightarrow \infty$ of the solution  (\ref{Uk_infinity}). In this limit we have

\begin{eqnarray}
\label{limit_uk}
u_k^\ua(t)&\rightarrow&  \frac{C}{\sqrt{2\,a(t)}}\left[-Q_\sca^2(k,t)\right]^{-1/2}\left\{ \frac{1}{2}\exp\left(-\int_{\upsilon_\sca}^{t}\left[-Q_\sca^2(k,t)\right]^{1/2} \D t\right)\right.\\
\nonumber
& +&\left.\im\,\exp\left(\int_{\upsilon_\sca}^{t}\left[-Q_\sca^2(k,t)\right]^{1/2} \D t\right)\right\},
\end{eqnarray}
where $C$ is a phase factor. Notice that  Eq.  (\ref{limit_uk})  is identical to Eq.  (\ref{ch4_uk_right}) obtained in the first-order phase-integral approximation.
Using Eqs. (\ref{ch4_a_fit}),  (\ref{ch4_zs}) and the growing part  of the solutions   (\ref{limit_uk}),  one can compute the scalar and power spectrum using the uniform approximation method,

\begin{eqnarray}
P_\sca(k)&=&\lim_{-k t\rightarrow \infty} \frac{k^3}{2\pi^2} \left|\frac{u_k^\ua(t)}{z_\sca(t)}\right|^2.
\end{eqnarray} 
We using the second-order improved uniform approximation for the power spectrum \cite{habib:2005b},

\begin{equation}
 \tilde{P}_{\sca}(k)=P_{\sca}(k)\left[\Gamma^*(\bar{\upsilon}_{\sca})\right],
\end{equation}
where $\bar{\upsilon}_{\sca}$ is the turning point for the scalar  power spectrum and 

\begin{equation}
\Gamma^*(\upsilon)\equiv 1+\frac{1}{12\upsilon}+\frac{1}{288\upsilon^2}-\frac{139}{51840\upsilon^3}+\cdots.
\end{equation}

%************************************************************************************
\subsection{Slow-roll approximation}
%************************************************************************************

The scalar  power spectra in the slow-roll approximation to second-order is given by the expression \cite{stewart:2001,gong:2004}

 \begin{eqnarray}
\label{ch4_sr_PS}
\nonumber
P_\sca^{\sr}(k)&\simeq&\left[1+(4b-2)\epsilon_1+2b\delta_1+ \left(4b^2 - 23+\frac{7\pi^2}{3}\right)\epsilon_1^2+\left(3b^2+2b-22+\frac{29\pi^2}{12}\right)\epsilon_1\delta_1 \right.\\
&+&\left.\left(3b^2-4+\frac{5\pi^2}{12}\right)\delta_1^2+\left(-b^2+\frac{\pi^2}{12}\right)\delta_2\right]\left.\left(\frac{H}{2\pi}\right)^2\left(\frac{H}{\dot{\phi}}\right)^2\right|_{k=aH} ,
\end{eqnarray}
where  $b \simeq 0.7296$ is the Euler constant. The spectral index in the slow-roll approximation  to second-order is

\begin{eqnarray}
\label{ch4_sr_nS}
n_\sca^{\sr}(k)&\simeq&1-4\epsilon_1-2\delta_1+(8c-8)\epsilon_1^2+(10c-6)\epsilon_1\delta_1 .
\end{eqnarray}

The parameters used in the equations (\ref{ch4_sr_PS}) and (\ref{ch4_sr_nS}), express in terms of the potential $V(\phi)$, are given by \cite{stewart:2001}

\begin{eqnarray}
\label{H2}
H^2 &\simeq&\frac{V}{3} \left(1+\frac{1}{6} U_1-\frac{1}{12} U_1^2+\frac{1}{9} U_1 V_1 \right),\\
\left(\frac{H}{\dot{\phi}}\right)^2&\simeq& \frac{1}{U1}\left(1+\frac{2}{3} U_1-\frac{2}{3} V_1-\frac{4}{9} U_1^2+\frac{7}{9} U_1 V_1-\frac{1}{9} V_1^2-\frac{2}{9} V_2\right),\\
\epsilon_1&\simeq&\frac{1}{2} U_1-\frac{1}{3} U_1^2+\frac{1}{3} U_1 V_1+\frac{4}{9} U_1^3-\frac{5}{6} U_1^2 V_1+\frac{5}{18} U_1 V_1^2+\frac{1}{9} U_1 V_2,\\
\delta_1&\simeq&\frac{1}{2} U_1-V_1-\frac{2}{3} U_1^2+\frac{4}{3} U_1 V_1-\frac{1}{3} V_1^2-\frac{1}{3} V_2+\frac{3}{2} U_1^3-4 U_1^2 V_1+\frac{23}{9} U_1 V_1^2\\
\nonumber
&-&\frac{2}{9} V_1^3 +\frac{17}{18} U_1 V_2-\frac{2}{3} V_1 V_2-\frac{1}{9} V_3,\\
\delta_2&\simeq&U_1^2-\frac{5}{2} U_1 V_1+V_1^2+V_2-\frac{19}{6} U_1^3+\frac{55}{6} U_1^2 V_1-\frac{13}{2} U_1 V_1^2+\frac{2}{3} V_1^3\\
\nonumber
&-&\frac{5}{2} U_1 V_2+2 V_1 V_2+ \frac{1}{3} V_3.
\end{eqnarray}
With,

\begin{eqnarray}
U_1&\equiv&\left(\frac{V'}{V}\right)^2 ,\\
V_1&\equiv& \frac{V''}{V},\\
V_2&\equiv&\frac{V'V'''}{V^2},\\
\label{V3}
V_3&\equiv&\frac{\left(V'\right)^2 V''''}{V^3}.
\end{eqnarray}

\bigskip
The expressions  (\ref{H2})-(\ref{V3}) depend explicitly on $\phi$. In order to compute the scalar  power spectrum we need to obtain the dependence  on the variable $k$. For a given value of $k$ ($0.0001 \,\Mpc^{-1} \leq k \leq 15 \,\Mpc^{-1}$) we obtain  $t$ from the relation $k=aH$. Thus, for each  $k$ one obtains a value of  $t$ that  we substitute into  Eq.  (\ref{phisr}).

%************************************************************************************
\subsection{Numerical solution}
\begin{figure}[th!]
\begin{center}
\vspace{1.5cm}
\includegraphics[scale=0.50]{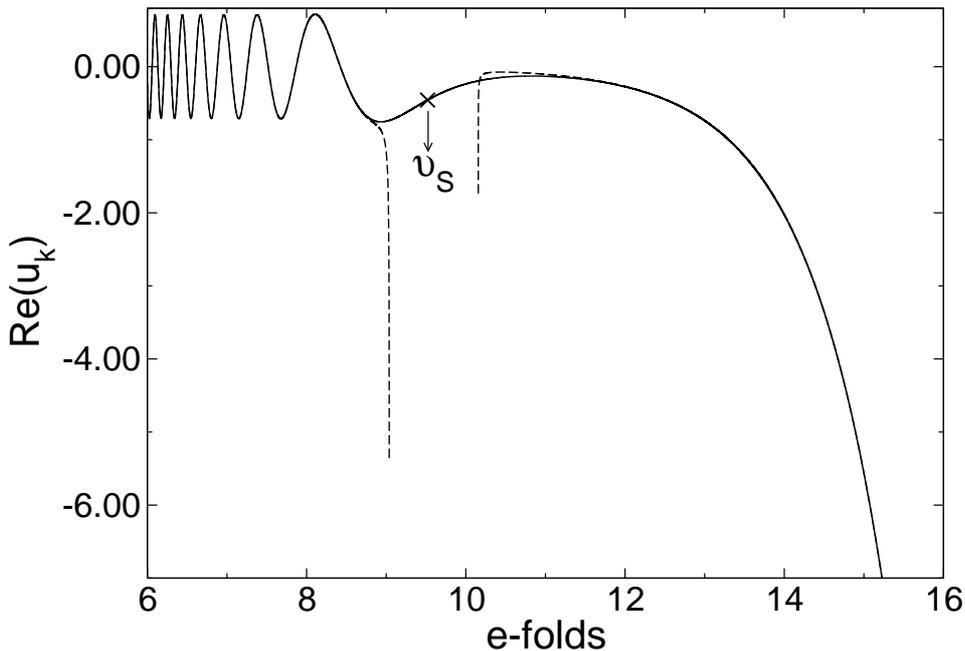}\\
\bigskip
\caption{$\real(u_k)$  versus the number of  e-folds for the chaotic inflationary  $\frac{1}{4}\lambda\phi^4$ model.  Solid line: numerical result (ICs phi3); dashed line: fifth-order phase-integral approximation.}
\label{ch4:Re(uk)}
\end{center}
\end{figure}

%************************************************************************************
We integrate on the physical time $t$ the equations (\ref{ch4_ddotUk}) governing the scalar  perturbations  using the predictor-corrector Adams method of order $12$ \cite{gerald:1984}, and  solve two differential equations, one for the real part and another for the imaginary  part $U_k$.  Two initial conditions are needed in each case  $U_k(t_i)$,  $U'_k(t_i)$, which can be obtained from the third-order phase-integral approximation.  We start the numerical integration at $t_i$ calculated at $25$ oscillations before  reaching the turning  point $t_\ret$ \cite{cunha:2005}. We call this procedure  ICs phi3. Fig. \ref{ch4:Re(uk)} compare the numerical solution with the fifth-order phase-integral approximation  for $\real(u_k)$. Figures are plotted against the number of e-folds   $N$.  The solid line corresponds to the numerical solution (ICs phi3), the dashed line  corresponds to the fifth-order phase-integral approximation. The turning point $\upsilon_\sca$ is indicated with an arrow. We stop the numerical computation of  $P_\sca(k)$ at $t= 5.00\times10^5\,M_\Pl^{-1}$, after  the mode leaves the horizon, where  $u_k/z_\sca$ is approximately constant. Notice that the expressions for  fitting  (\ref{ch4_a_fit}) and (\ref{ch4_phi_fit}) are valid in the aforementioned  time scales, therefore we can use them for computing  the scalar $P_\sca(k)$  power spectrum.

%************************************************************************************
\section{Results}
\label{results}
%************************************************************************************

\begin{figure}[th!]
\vspace{1.5cm}
\begin{center}
\includegraphics[scale=0.50]{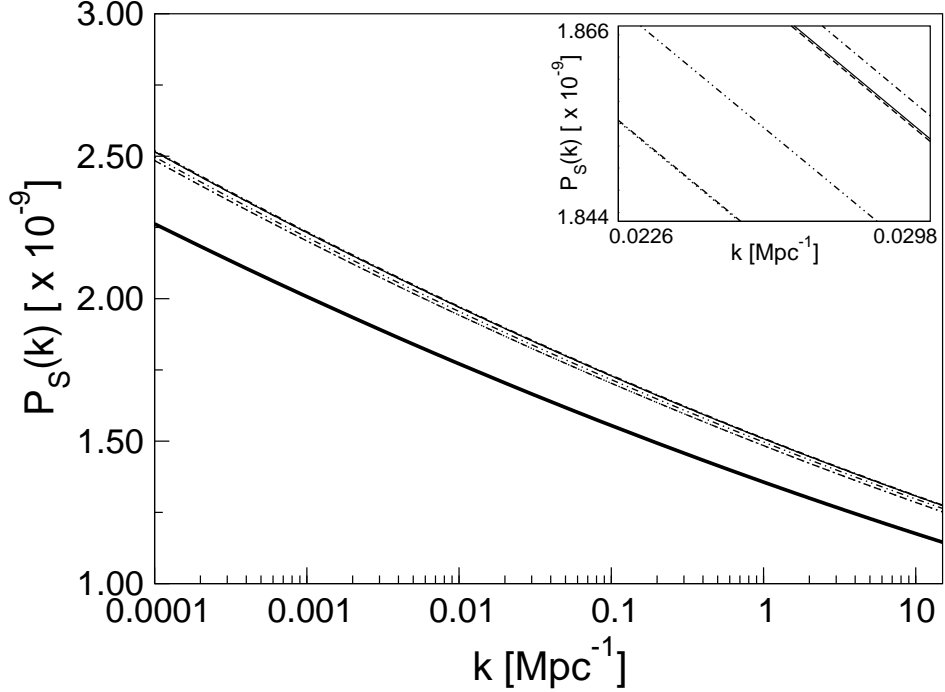}
\caption{\small{$P_\sca(k)$  for the chaotic inflationary $\frac{1}{4}\lambda\phi^4$ model.  Thin solid line: numerical result (ICs phi3); dot-dashed line: third-order phase-integral approximation; dashed line: fifth-order phase-integral approximation; thick solid line: first-order phase-integral approximation, WKB and  first-order uniform approximation; dashed double-dots line: second-order improved uniform approximation, double-dashed dot line: second-order slow-roll approximation; dotted line: first-order slow-roll approximation. The inset is an enlargement of the figure.}}
\label{ch4:PS}
\end{center}
\end{figure}

For the chaotic $\frac{1}{4}\lambda\phi^4$ inflationary model,  we want to compare the scalar power spectra for different values of  $k$ calculated using the third and fifth-order  phase-integral approximation  with the numerical result (ICs phi3), the first and second-order slow-roll approximation and  the first and second-order uniform approximation method.  First we analyze the results for the scalar $P_\sca(k)$  power spectrum  shown in Fig. \ref{ch4:PS}. 

%*****************

Table \ref{ch4_t1} shows the value of $P_\sca(k)$ and the relative error $P_\sca(k)$ using each method of approximation at the WMAP pivot scale \cite{habib:2005b}. It can be observed that the best value is obtained with the fifth-order phase-integral approximation. It should be noticed that the   slow-roll approximation works well since  the parameters $\epsilon_1$, $\epsilon_2$ and $\delta_n$ are small. 

\begin{table}
{\begin{tabular}{l|c|c}
\hline\hline
\raisebox{-0.5ex}{\hspace{3.5cm} Method} &\raisebox{-0.5ex}{$P_\sca(k)\times 10^{-9}$}  & \raisebox{-0.5ex}{$|$rel. error $P_\sca(k)$$|$}\\
\hline\hline
Numerical                                                            & $1.7995$ &                     \\
Third-order phase-integral approximation            & $1.8021$ & $0.1407\, \%$\\
Fifth-order phase-integral approximation             & $1.7993$ & $0.0149\, \%$\\
Leading order of slow-roll approximation             & $1.7889$ & $0.5866\, \%$ \\
First-order slow-roll approximation                      & $1.7952$ & $0.2394\, \%$ \\
Second-order slow-roll approximation                  & $1.7953$ &  $0.2324\, \%$\\
First-order phase integral approximation, WKB,   & $1.6181$ & $10.0851\, \%$\\
and first-order uniform approximation                 &                &                         \\
Second-order improved uniform approximation   & $1.7857$ & $0.7694\, \%$   \\
\hline\hline
\end{tabular}}
\caption{Value of $P_\sca(k)$ and its relative error obtained  with different approximation methods for the chaotic inflationary model  $\frac{1}{4}\lambda\phi^4$ for the mode  $k=0.05\,\Mpc^{-1}$.}
\label{ch4_t1}
\end{table}

\begin{figure}[htbp]
\begin{center}
\vspace{1.4cm}
\includegraphics[scale=0.50]{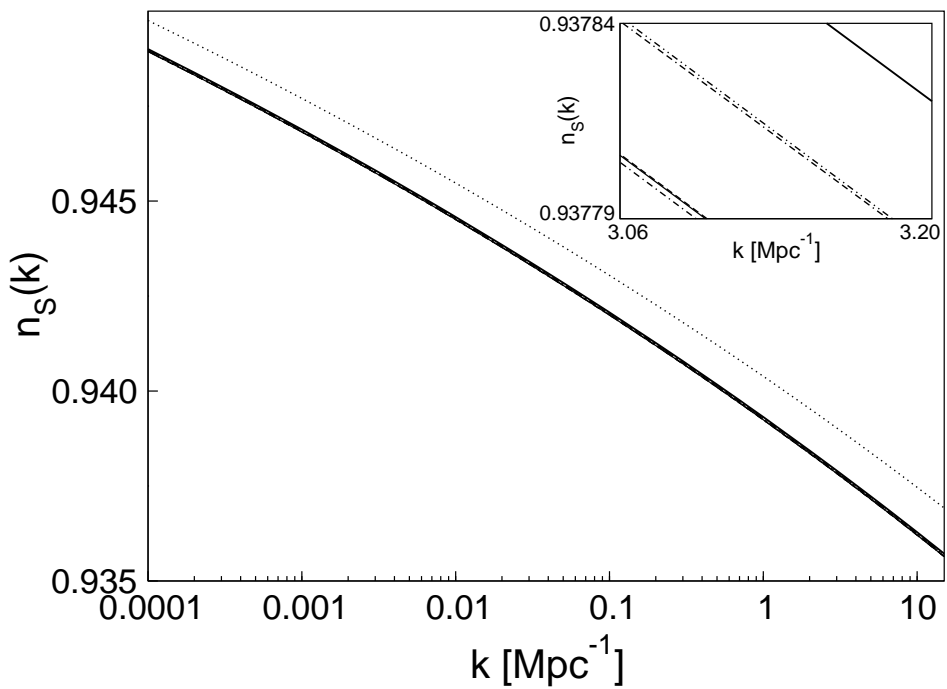}
\caption{$n_\sca(k)$  for the chaotic inflationary $\frac{1}{4}\lambda\phi^4$ model.  Thin solid line: numerical result (ICs phi3) and fifth-order phase-integral approximation; dot-dashed line: third-order phase-integral approximation; thick solid line: first-order phase-integral approximation, WKB and  first-order uniform approximation; dashed double-dots line: second-order improved uniform approximation; double-dashed dot line: second-order slow-roll approximation; dotted line: first-order slow-roll approximation. The inset is an enlargement of the figure.}
\label{ch4:nS}
\end{center}
\end{figure}

The first-order phase integral approximation, the WKB and the first-order uniform approximation give the same result, and deviate from the numerical result in  $10.0851\,\%$. The second-order-improved uniform approximation gives an error of $0.7694\,\%$. With the leading, first and second-order slow-roll approximation we have an error for $P_\sca(k)$   of $\gg 100\,\%$, $0.2394\,\%$ and $0.2324\,\%$, respectively.  Using the third-order phase-integral approximation the error gives $0.1407\,\%$, whereas the fifth-order phase-integral, due to the convergence of the method, reduces to  $ 0.0149\,\%$  for $P_\sca(k)$. Note that rel. error $P_\sca(k)$ with sr1 $\approx$ rel. error $P_\sca(k)$ with sr2, which is unexpected according to the results obtained previously for the power-law inflationary model \cite{habib:2004}. The computation time is different for each method, for the scalar power spectrum we have the following times: numerical $\approx$ several hours, pi3  $\approx$ few minutes, pi5 $\approx$ several minutes, sr0 $\approx$ few seconds, sr1 $\approx$ few seconds, sr2 $\approx$ few seconds, WKB, ua1, pi1 $\approx$  few seconds, ua2 $\approx$  few seconds.
Fig. \ref{ch4:nS}  shows the results for the  spectral indice $n_\sca(k)$.

%*****************

Table \ref{ch4_t2} shows the value of $n_\sca(k)$ and the relative error $ n_\sca(k)$ using each method of approximation at the WMAP pivot scale \cite{habib:2005b}.  The first-order phase integral approximation, the WKB and the first-order uniform approximation give the same result, and deviate from the numerical result in  $0.007239\,\%$. The second-order-improved uniform approximation gives an error of $0.003909\,\%$. With the leading, first and second-order slow-roll approximation we have an error for $n_\sca(k)$ of $6.0692239\,\%$, $0.116637\,\%$ and $0.0012000\,\%$, respectively.  Using the third-order phase-integral approximation the error gives $0.000157\,\%$, whereas the fifth-order phase-integral, due to the convergence of the method, reduces to  $0.000026\,\%$  for $n_\sca(k)$. Note that the first and second-order slow-roll approximation gives a better results for the  spectral indice $n_\sca(k)$ than the power spectrum $P_\sca(k)$.

\vspace{0.5cm}
\begin{table}[htbp]
{\begin{tabular}{l|c|c}
\hline\hline
\raisebox{-0.5ex}{\hspace{3.5cm} Method} &\raisebox{-0.5ex}{$n_\sca(k)$}  & \raisebox{-0.5ex}{$|$rel. error $n_\sca(k)$$|$}\\
\hline\hline
Numerical                                                            & $0.942780$ &                     \\
Third-order phase-integral approximation            & $0.942779$ & $0.000157\, \%$\\
Fifth-order phase-integral approximation             & $0.942781$ & $0.000026\, \%$\\
Leading order of slow-roll approximation             & $1$               & $6.069239\, \%$ \\
First-order slow-roll approximation                      & $0.943880$ & $0.116637\, \%$ \\
Second-order slow-roll approximation                  & $0.942894$ &  $0.012000\, \%$\\
First-order phase integral approximation, WKB,   & $0.942849$ & $0.007239\, \%$\\
and first-order uniform approximation                 &                      &                        \\
Second-order improved uniform approximation   & $0.942817$ & $0.003909\, \%$   \\
\hline\hline
\end{tabular}}
\caption{Value of $n_\sca(k)$ and its relative error obtained  with different approximation methods for the chaotic inflationary model  $\frac{1}{4}\lambda\phi^4$ for the mode  $k=0.05\,\Mpc^{-1}$.}
\label{ch4_t2}
\end{table}

\bigskip
The relative errors with respect to the numerical result have been obtained using the expressions:

\begin{eqnarray}
\text{rel. error}\,\,P_{\sca}(k) &=& \frac{\left[P_{\sca}^\mathrm{approx}(k)-P_{\sca}^\num(k)\right]}{P_{\sca}^\num(k)}\times 100,\\
\text{rel. error}\,\,n_{\sca}(k) &=& \frac{\left[n_{\sca}^\mathrm{approx}(k)-n_{\sca}^\num(k)\right]}{n_{\sca}^\num(k)}\times 100.
\end{eqnarray}

%************************************************************************************
\section{Conclusion}
\label{conclusion}
%************************************************************************************

As in our previous work \cite{rojas:2007b,rojas:2007c,rojas:2009},  we show that the phase-integral approximation is an efficient method to calculate the scalar (or tensor) power spectrum and the spectral index in several models of inflation. 
Our results indicate that the scalar power spectrum is almost scale invariant since we have considered the usual scenario where the observable modes cross the horizon before the inflation ends,  unlike the proposal presented by Ramirez {\it et. al} \cite{ramirez:2009}. Therefore, the good agreement between the numerical results and those obtained with the phase-integral approximation shows that the phase integral method is a very useful approximation tool for computing the scalar and tensor the power spectra in  a wide range of inflationar scenarios.

%-------------------------------------------------------------------------------------

%\bibliographystyle{JHEP}
%\bibliography{phi4}

\end{document}